\documentclass[structabstract]{aa}  
\usepackage[utf8]{inputenc}

\usepackage{graphicx}
\usepackage{natbib}
\usepackage{longtable,lscape}
\usepackage{rotating}

\usepackage{txfonts}
\newcommand{\lsun}{\mbox{$\rm L_\odot\,$}}

\def\arcsec{\hbox{$^{\prime\prime}$}}

\begin{document}
   \title{The ionizing sources of luminous compact HII regions in the RCW106 and RCW122 clouds}

   \author{J. M. C. Grave\inst{1,3}
          \and
           M. S. N. Kumar\inst{2,3}
           \and
	   D. K. Ojha\inst{4}
           \and
          G. D. C. Teixeira\inst{3}
	  \and 
          G. Pace\inst{3} }

   \institute{ FCNET, Universidade Lusófona do Porto, 
	      Rua Augusto Rosa, nº 24, 4000-098 Porto, Portugal\\
              \email{nanda@astro.up.pt}
         \and
            Rua Ruben A, 137, 10 Esq, 4150-762 Porto, Portugal
         \and
	    Centro de Astrofísica da Universidade do Porto,
              Rua das estrelas, 4150-762 Porto, Portugal  
         \and
	      Tata Institute for Fundamental Research, Homi Bhabha Road, Mumbai 400005, India}


 \abstract{}{}{}{}{} 
 
\abstract {Given the rarity of young O star candidates, compact HII
  regions embedded in dense molecular cores continue to serve as
  potential sites for peering into the details of high-mass star
  formation.}{We uncover the ionizing sources of the most luminous and
  compact HII regions embedded in the RCW106 and RCW122 giant
  molecular clouds, known to be relatively nearby (2--4 kpc) and
  isolated, thus providing an opportunity to examine spatial scales of
  a few hundred to a thousand AU in size.} {High spatial resolution
  (0.3\arcsec), mid-infrared spectra (R=350), including the fine
  structure lines [ArIII] and [NeII], were obtained for four luminous
  compact HII regions embedded inside the dense cores within the
  RCW106 and RCW122 molecular cloud complexes. At this resolution,
  these targets reveal point-like sources surrounded by the nebulosity of
  different morphologies, thereby uncovering details at spatial dimensions of
  $<$1000\,AU. The point-like sources display [ArIII] and [NeII] lines
  -- the ratios of which are used to estimate the effective
  temperature of the embedded sources.}  { The derived temperatures
  are indicative of mid-late O type objects for all the sources with
  [ArIII] emission.  Previously known characteristics of these targets
  from the literature, including evidence of disk or accretion, 
  suggest that the identified sources may grow more to become early-type O 
  stars by the end of the star formation process.}{}



   \maketitle
%

\section{Introduction}

Massive stars play a determining role in shaping the structure and
evolution of the Galaxy; however, it seems almost surprising that our
understanding of their formation and early evolution, in particular,
has barely begun. Forming inside cocoons of dense gas and dust which
are not easily probed at short wavelengths, the most well studied
early manifestation of a young massive star is the appearance of a
compact or ultracompact (UC) HII region, usually detected at radio
wavelengths. However, the high column densities associated with these
objects prevent us from observing the accreting object directly at
this critical early stage, at least until the extinction falls below a
few magnitudes in the K band. Furthermore, OB stars are known to form
in clusters or multiple systems, which when associated with the large
distances where they are usually found, makes it difficult to probe
them as isolated entities.

Until recently, only a few objects that can be called O-type young
stars have been observationally identified in UCHII regions
\citep{watson97,MHernandez03}, and a large number of the so-called
\textquotedblleft high mass protostellar
candidates \textquotedblright are known to either host accreting
B-type stars \citep{grave09}, or, are doubted to be clusters. Thus, strong
efforts are being made to remove contaminants and to identify a usable
sample of massive young stellar objects for follow-up observations in
order to examine the detailed mechanism of high-mass star formation
\citep{mottram07}.  After the signatures of accretion are uncovered in
luminous far-infrared targets, establishing the proper content and
temperature of the embedded population is crucial for identifying
whether the accreting object is indeed an O-type star. Uncovering the
embedded star content is possible with high spatial resolution
observations \citep[e.g.][]{feldt03}, provided the distance to the
target is well known and preferably resolved. Only then can pointed
observations be undertaken at similar spatial resolutions to evaluate
the temperature or mass associated with the candidate high-mass stars.

In this context, the technical evolution in mid-infrared (MIR)
observations has provided a useful tool for peering into obscured
regions, with a spatial resolution of a few hundred to a thousand AU
for targets located closer than 4 -- 5 kpc. A number of fine-structure
lines produced in the ionized regions surrounding massive stars can be
observed at these wavelengths, and their analysis can be used to try
to uncover the properties of ionizing stars. This method is analogous
to what is done with forbidden lines in the optical for more evolved
stars. However, they are undetectable in these embedded regions. In
some cases, the properties of the ionizing star have been derived
using helium and hydrogen lines, which are detected in the near
infrared \citep{hanson02}. The line intensities and their ratios
reflect the ionization status of the elements in UC HII
regions. Detecting even these near-infrared lines is quite challenging
in O star-forming regions that are highly extincted and obscured. But
they lend themselves to being probed at wavelengths longwards of a few
microns where emission from elements such as argon and neon are
found. The ionization potentials to form Ne$^+$ and Ar$^{2+}$ are
21.56 and 27.63 eV, respectively.

These are higher values than are required for H$^+$ or He$^+$, so the
line ratios should be more sensitive to the characteristics of
ionizing objects with early spectral types. However, when performing
this type of analysis, one should also take the influence of the
nebular contamination into account and, especially, the influence of
neighbouring ionizing stars on the line intensities, before arriving
at any conclusions regarding the nature of an individual massive young
star. For example, fine-structure lines have previously been observed
in the MIR N-band \citep{fujiyoshi98,fuji01} and the far-infrared
(FIR) \citep{okada10}, in an effort to uncover the properties of the
ionizing sources in high-mass star-forming regions. However, the
observed sources were the combined output of compact clusters, and the
results did not lead to a unique interpretation of the nature of the
ionizing sources in those clusters.

As such, one factor that can significantly increase the reliability of
the results obtained with this method is the spatial resolution with
which the observations are done. The diffraction-limited observations
with 8-m class telescopes in the MIR band, resulting in a resolution
of $\sim$0.3\arcsec\,, allows us to probe the UC HII regions, sampling
typical dimensions from a few hundred up to a thousand AU for targets
located within 2-4kpc of the Sun.

In this work, we present such observations for four high-luminosity
sources ($>$10$^4$\lsun), which are located in nearby (2--4 kpc) star-forming 
complexes, RCW106 and RCW122, taking advantage of the
diffraction limit achieved at the Very Large Telescope (VLT) from the
European Southern Observatory (ESO).

\section{The sample}

The targets RCW106 and RCW122, located at distances of 3.6$\pm0.6$ kpc
\citep{lockman79} and 2.7$\pm0.5$ kpc \citep{russ03}, respective ,
were identified as HII regions by their H-${\alpha}$ emission
\citep{rcw60} and are known to be associated with giant molecular
cloud complexes \citep[e.g.][and references
  therein]{lo09,mook04,arnal08}. These clouds are associated with some
of the densest and most compact cores harbouring very luminous FIR
sources and compact HII regions. RCW106-MMS5 and RCW106-MMS68 stand
out as the most massive and densest cores found in the entire RCW106
cloud \citep{mook04} and display intense emission from a large number
of molecular species \citep{lo09}. Both these cores are associated
with extraordinarily luminous FIR sources ($>$10$^5$\lsun). Similarly,
the cores chosen in RCW122 stand out as the most luminous FIR sources
($>$10$^4$ \lsun), coinciding with compact HII regions and
representing O-type stars \citep{ghosh89}. The RCW106-MMS5 core
coincides with the G333.6-0.22 compact HII region known to emit Lyman
continuum flux density that can only be achieved by a cluster of
several late O type stars \citep{fuji06}. Furthermore, the
near-infrared images analysed by the same authors show that the
ionizing sources in this region are represented by unresolved
point-like objects at a level of 1\arcsec\,. In Table
\ref{table:sample}, we summarize the basic details of the chosen
sample.

\section{Observations and data reduction}

\begin{figure}
   \centering
   \includegraphics[width=9cm]{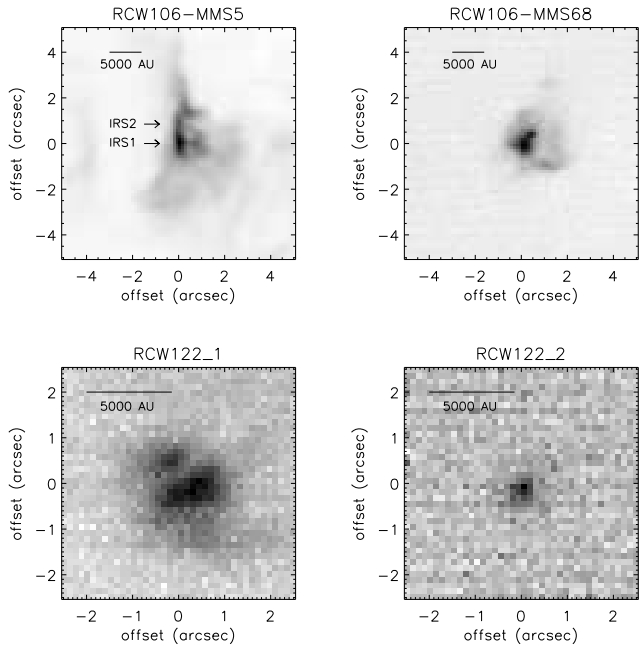}
      \caption{N band aquisition images of the observed targets for
        which spectra were obtained. The scale bars indicate a
        physical size of 5000 AU, assuming distances of 3.6\,kpc and
        2.7\,kpc to RCW106 and RCW122, respectively. The
          offsets marked in the figure are with respect to the
          position in Table\,1, which correspond to the peak emission
          in each image, except for RCW\,122\_1 where it represents
          the midpoint of the dark lane separating the two emission
          peaks.}
         \label{images}
   \end{figure}

The observations were carried out in visitor mode with the MIR VISIR
camera, mounted on the Cassegrain focus of the 8m VLT UT3 during the
nights of 2-3 June 2005. Low-resolution (R=350 at 10$\mu$m) N-band
spectra and the respective acquisition images were obtained for each
of the targets listed in Table \ref{table:sample}. The sky conditions
were excellent, and the system yielded diffraction-limited observations
at a spatial resolution of 0.3\arcsec\,. The slit length and width
used are 32.3\arcsec\ and 0.75\arcsec, respectively. Four spectral
settings overlapping by, at least, 15$\%$ were used to cover the full
N band (central wavelengths = 8.8, 9.8, 11.4, and 12.4 $\mu$m). The
total on-source integration time was two minutes for each spectroscopic
setting. To avoid contamination from the extended diffuse emission,
judged by the relatively deep acquisition images with eight seconds of
integration time, a chopper throw of 10\arcsec\,, and nodding on slit
were applied. Early-type stars were observed before and after each
target observation to obtain flux calibration (HD146791, HD184996, and
HD171759). In the particular case of RCW106-MMS5, the slit orientation
was chosen so that it contained the two bright sources simultaneously
(IRS1 and IRS2 in Fig. \ref{images}).

The VISIR pipeline, IRAF routines, and customized IDL scripts were used
for data reduction. We used the pipeline to perform the subtraction of
the chopping and nodding pairs in order to remove most of the sky and
telescope background. Any residual background contribution was removed
while extracting the spectra by fitting the background emission at
each row in the spectral image and subtracting it. The resulting
images were shifted and co-added manually to maximize the resulting
signal-to-noise ratio and spatial resolution. Wavelength calibration
and removal of the curvature in the spectroscopic data were done by
tracing sky emission lines. Spectra were extracted in different
apertures. In the case of RCW106-MMS5, we used smaller apertures
than for the other sources in order to extract the spectra of the two
MIR sources (IRS1 and IRS2 in Fig.\,1). These sources are separated by
a distance of 1\arcsec\,, both embedded in an extended bright
nebula. Thus, the extracted spectra may still carry some
contamination. Absolute flux calibration was obtained by normalizing
the standard star spectra to VISIR narrowband fluxes. In the MIR
ground-based observations, it is quite challenging to obtain an
accurate flux calibration. We analysed the variability in the
sensitivity of the standard star observations throughout the observing
run, which lasted five hours, and we noticed a maximum amplitude that
yields an accuracy of roughly 50 \%. Considering that we only
calibrate each science target observation with the adjacent standard
star observations, the accuracy for the flux calibration of each
object spectrum is 30\%.

\begin{table}
\caption{Details of the observed sample.}             
\label{table:sample}      
\centering                          
\begin{tabular}{c c c c c c}        
\hline\hline               
Name & RA & DEC & d & L \\
 & (J2000) & (J2000) & (kpc) & ($10^4$\lsun) \\
\hline                        
   RCW106-MMS5 & 16:22:09.6 & -50 06 00 & 3.6 &  $<90$\\      
   RCW106-MMS68 & 16:20:11.2 & –50:53:15.8 & 3.6 & $25$\\
   RCW122\_1 & 17:20:07.9 & -38:57:15 & 2.7 &  $<2.4$\\
   RCW122\_2 & 17:20:06.9 & -38:56:59 & 2.7 & $<2.4$\\
\hline                                   
\end{tabular}
\end{table}

\section{Results}

In Fig. 1, we display the acquisition images of the four observed
targets. It can be seen from these images that several details of the
observed MIR peaks enclosed within projected diameters of 5000\,AU
begin to be resolved. We note that the peaks in the RCW106 sources
appear point-like in the near-infrared adaptive optics imaging data
obtained with the NACO instrument on VLT \citep{kumar13}. In
particular, the source IRS2 marked on the RCW106-MMS5 image is a well
identified point source at a spatial resolution of 0.l\arcsec\ in the
near-infrared, likely representing an individual star unlike IRS1
which peaks between 8--18$\mu$m. At 0.3\arcsec resolution, RCW122\_2
and the peak of RCW106-MMS68 appear point-like (Fig.\,1) while
RCW122\_1 reveals a dark lane separated by a compact bipolar
nebula. The projected sizes are typically less than 5000\,AU (see
  Fig.1 for the scale bar). Therefore, the MIR peaks shown in Fig. 1
are in all likelihood isolated objects of high infrared luminosity and
not resolved clusters.

\begin{figure}
   \centering
   \includegraphics[width=9cm]{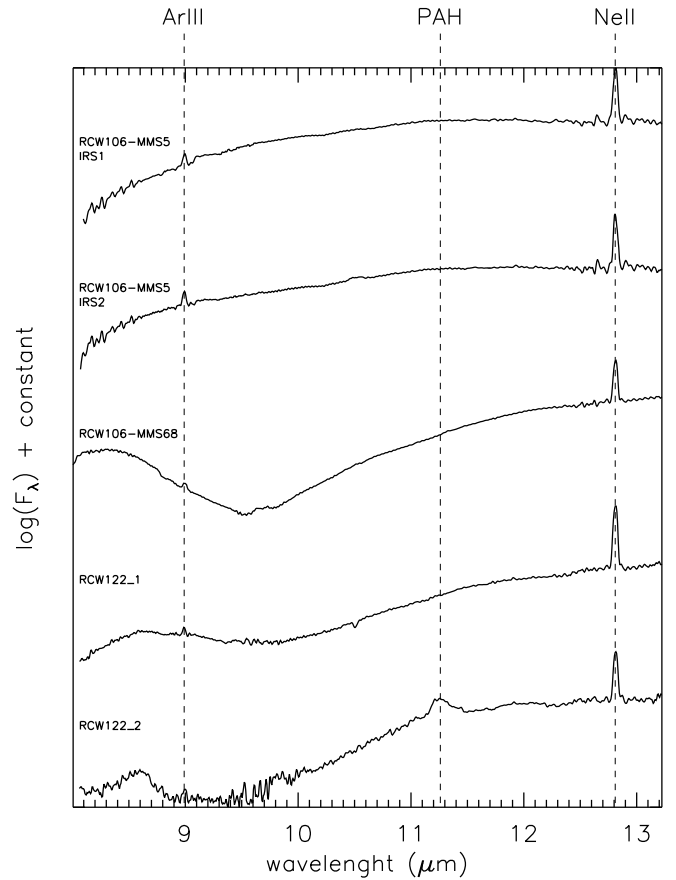}
      \caption{N-band spectra of the observed sources. The dashed lines
        mark the most prominent spectral features detected.}
         \label{spectra}
   \end{figure}

In Fig. \ref{spectra} we present the spectra obtained with VISIR in
the N band for the five detected sources in our sample, consisting of
four observations. These are wavelength- and flux-calibrated, and the
absolute fluxes are shifted by an arbitrary constant for convenience
of plotting in a single figure. The [NeII] line at 12.8 $\mu$m is
detected in all the sources, while the [ArIII] at 9 $\mu$m is also
visible in all but RCW122\_2. On the other hand, only this source
displays the PAH emission feature clearly at $\sim$11.3 $\mu$m. The
[SIV] line at 10.5 $\mu$m typically observed in these type of objects
is not detected in these sources, although there is a shallow absorption
feature in the spectrum of RCW122\_1, which we cannot associate with
that line. Recent studies \citep{furn10} have shown that the
[NeII]/[SIV] ratio is a better diagnostic for obtaining the
temperature. In the absence of the [SIV] line, we used the
[ArIII]/[NeII] diagnostic, taking the uncertainties
pointed out by \citet{furn10} into account.


Examination of the spectral images shows that for RCW106-MMS5,
the emission of both lines clearly peak on IRS1 and IRS2 and extends
at a low level to fill the region between them. Given that the
separation between those sources is about 3600\,AU, the low-level
emission in between the sources may be the combined contribution of
both. In RCW106-MMS68, the intensity of the lines is clearly
concentrated on the target -- a star visible in the near-infrared -- with the
centroid of the line emission well aligned with the centroid of the
continuum strip. In the RCW122 sources, the line emission is also
found to be concentrated on the actual targets. However, there
are some minor artefacts in the co-added final spectral image,
suggesting possible external contamination from beyond the sources.

The line fluxes were calculated for each of the observed lines and
corrected for the extinction. Following the work of \cite{furn10}, we
used the extinction-corrected line ratio between the [ArIII] and
[NeII] lines to obtain the temperature of the observed sources in each
region. It was computed using the relation
\begin{equation}
 (T_{eff}/10^3 K)_{2Z_{\odot}}=+48.92+12.92\log\left(\frac{I[ArIII]}{I[NeII]}\right)
\end{equation}
derived in the same work for regions with twice the metallicity of the
Sun. Given that both RCW106 and RCW122 are located in the
  fourth Galactic quadrant, with relatively well estimated distances,
  they should lie closer to the metal-rich Galactic centre, justifying
  the choice above.

\subsection{Extinction and related uncertainties}

The value of the extinction used to correct the line fluxes has a
significant impact on the calculated temperatures. It is difficult to
obtain an accurate extinction for these kinds of sources because of
numerous uncertainties. The molecular cores hosting these objects are
so dense that any attempt to measure extinction using light from
background stars is impossible. Measurements using (sub)mm dust
continuum emission to calculate column densities often lead to high
values of extinction. However, the uncertainties about the dust
properties and the degree to which the core may present a clumpy
structure may result in large variations of the extinction within a
narrow spatial range. Such variations should be averaged out by the
large beam of the (sub)mm observations. Therefore, this must be taken
into account when using the submm derived extinction values for the
analysis of objects with sizes of only a few thousand AU, such as the
ones studied in this work. Only RCW122 is associated with a resolved
cluster, where an infrared analysis can lead to an average
extinction, but it has not been done so far. 

Strong controversy
about measuring extinction at MIR wavelengths, which
affects the lines and the continuum emission to different extents
\citep[e.g.][and references therein]{jiang06}. Working within these
limitations and without ignoring any measurements, we have adopted
both the lowest and highest values of extinction and tabulated the
associated temperatures in Table \ref{tab:fluxes}. The higher values
in column 6 of this table are obtained using the uniform dataset from
the ATLASGAL survey \citep{schuller09}. The targets are assumed to be
located at the centre of the dense cores traced by 870 $\mu$m dust
continuum observations. The peak flux of the associated cores obtained
from the ATLASGAL database was used to compute the column
densities. Following \citet{draine03}, the extinction for each source
was calculated using the relation
A$_V$=N$_{H2}\times$6.71$\times$10$^{-22}$. Considering the
  dense and dusty cores associated with the sources analysed in this
  work, we have adopted $R_V=5.5$, since it corresponds to an
  extinction law that is more representative of the denser regions of
  molecular clouds, as shown for other, similar regions
  \citep{chapman09}.




\begin{table*}
 \centering
 \caption{Table of observed fluxes, extinctions, and effective temperatures derived using the equation of \cite{furn10}. The assumed uncertainty for the fluxes listed was 30$\%$.}
 \label{tab:fluxes}
\tabcolsep=0.11cm
\begin{tabular}{lcccccccc}

 \hline\hline\
Source & F$_{ArIII}$ & F$_{NeII}$ & A$_{V}$(low)\tablefootmark{a} & T (low) & A$_V$(high) & T(high) & SpT \\
 & (erg s$^{-1}$ cm$^{-2}$) & (erg s$^{-1}$ cm$^{-2}$) &  mag & $10^3$K & mag & $10^3$K & adopted\tablefootmark{b} \\
\hline
RCW106-MMS5 IRS1 & 2.6E-11 & 1.8E-10 & 16 & 41$\pm$7  & 119 & 59$\pm$8 & O5\\
RCW106-MMS5 IRS2 & 2.5E-11 & 9.6E-11 & 16 & 44$\pm$7  & 119 & 62$\pm$9 & O4\\
RCW106-MMS68     & 2.1E-12 & 1.3E-11 & 13 & 26$\pm$10 & 108 & 43$\pm$7 & O9 \\
RCW122\_1        & 5.9E-14 & 3.4E-12 & 20 & 29$\pm$10 & 79  & 40$\pm$8 & O8.5\\
RCW122\_2        & ...     & 7.1E-13 &... & ...    & ... & ... & ... \\
\hline
\end{tabular}

\tablefoottext{a}{Extinction estimates used for RCW106-MMS5, RCWMMS-68, and RCW122 are from \citet{fuji06}, \citet{bik06}, and \citet{arnal08} respectively.}

\tablefoottext{b}{see discussion section for the reasoning}

\end{table*}

\section{Discussion and conclusions}

As shown in the previous section, the 0.3\arcsec\ diffraction
  limit of the VLT MIR observations provides a spatial resolution
  that represents physical scales below 1000AU for all four observed
  targets, uncovering several details of the MIR peaks. The
  temperatures estimated by analysing nebular lines from these peaks
  are indicative of hot O type stars. In fact, \citet{hanson02} point out
  that for those objects with both NIR and MIR data, the temperature
  estimated from MIR nebular line analysis are systematically lower
  than photospheric temperatures. The issue here is whether more
  than one O-type star can contribute to the measurements presented
  here on a physical scale of 1000\,AU. Theoretical arguments propose
  that some level of multiplicity is expected either due to N body
  interactions \citep{bbv03,bb05} or to the fragmentation of massive disks
  \citep{km06,kkm10}, which results in multiple objects. However, such
  mechanisms do not produce equal mass objects on such small physical
  scales, which can dominate the ionizing radiation within the MIR
  peaks. In fact, equally dominant sources are expected to be found at
  larger separations \citep[e.g.][]{krum09} of greater than a thousand
  AU, which would be resolved as independent MIR peaks through our
  observations, as is the case for RCW106MMS5. A study of binarity in
  massive stars \citep{sana12}, focussing on evolved stages, also
  supports this conjecture. Therefore, the results presented in this
  work describe the nature of the most dominant ionizing source,
  typically representing O-type stars, even though a low level of
  contamination from multiple sources is expected.

The temperature estimates of spatially resolved MIR peaks in this
work, such as sources IRS 1 and IRS 2 associated with RCW106MMS5, may
represent objects of different masses at different evolutionary
stages. An exact description of the true nature and physical
conditions around the embedded high-mass star is beyond the reach of
our observations because the photosphere is not traced. Even so, the
wealth of information on these luminous targets from previous
studies allows us to speculate about some characteristics. If these
sources were to represent zero-age-main-sequence stars, our results
could be extrapolated to spectral types \citep{vacca96}. In
contrast, if they are younger objects, similar to bloated pre-main
sequence objects \citep[e.g.][]{och11} , representing accreting stages,
it would imply that they will become hotter and more massive when star
formation is complete. Given that the chosen sources are all embedded
inside compact HII regions with association of dense molecular gas, we
can safely rule out that they represent evolved objects. Infrared and
radio observations on angular scales of an arcsecond are available for
the sources in the RCW106 cloud but not for the RCW122 sources. Each
of the targets are discussed in detail below.

RCW106-MMS5: The targets studied here coincide with the core of the
UCHII region associated with the MMS5 core. A H90$\alpha$ radio
recombination line study of this compact region, surrounding targets
IRS1 and IRS2, displays double-peaked profiles, indicating a complex
kinematic structure \citep{fuji06}. These authors modelled this region
to be associated with an outflow originating in a compact object that
represents a cluster of O type stars. These results, combined with the
the small spatial extent of the sources in the MIR clearly
excluding the possibility of clusters or multiple systems of ionizing
stars and with the information that only two stars (IRS1 and IRS2) are
found in this region, imply that the objects are at an early
evolutionary stage and, at least one of them is, in all likelihood,
are still accreting. The lower limit of temperature in Table~\ref{tab:fluxes} 
suggests that they are similar to O6 main-sequence
stars, while the upper limit would place them at the Wolf-Rayet
limits. Since these two are the most luminous point-like objects within
the compact HII region, producing a LyC flux equivalent to 12 O7 type
stars, the lower and upper limits both seem unreasonable. However, if
we assume that IRS1 and IRS2 are in the range of O3-O5 spectral types,
much of the observed data, in particular the bolometric luminosity and
the LyC flux, can be coherently explained.

RCW106-MMS68: This source was studied using VLT-ISAAC spectroscopy by
\citet{bik06} (object 16164nr3636), which led them to model the CO-band
head emission to arise in a disk. They also suggest that the
embedded source is of spectral type O9V to interpret the origin of the
nebular HeI emission. The extinction of 13 visual magnitudes, as
adopted by \citet{bik06}, is clearly an under-estimation of the real
value, as stated by the authors, because it was derived from neighbouring,
more evolved stars. The presence of the CO band heads and the deep 10-micron 
silicate absorption visible in the respective spectrum in
Fig. 1, indicates a higher degree of obscuration.

Spectro-astrometric analysis of the Br$\gamma$ line from this object
indicated the presence of an expanding HII region \citep{grave07} that
traces size scales of a few hundred AU. Considering the analysis of
\citet{bik06}, their estimated spectral type is well within the
temperature limits presented in this work. This is clearly an
excellent target for follow-up observations at other wavelengths and
comparable spatial resolution, to peer into the mechanisms of O star
formation.

RCW122\_1 \& RCW122\_2: These objects are the two main MIR peaks
located at the heart of the largest clump traced by radio and FIR
observations. The two objects (1 \& 2) are separated by
$\sim$17\arcsec\ and are identified as RCW122 A \& B by
\citet{ghosh89}. This region appears as an elongated object with a
single peak on the 870$\mu$m images obtained through ATLASGAL. The
molecular clump clearly hosts an embedded cluster, which is traced by resolved
stellar population visible in the 2MASS and Spitzer-IRAC
data. RCW122\_1 reveals itself to be a spatially extended object,
separated by the dark lane representative of an edge-on disk like
object. The true driving engine may actually be hidden, and it is
possible that our observations trace the scattered light within
outflow cavities. Even so, the computed value for the temperature is
high, suggesting a late O-type object. We think that RCW122\_1 is of
the spectral type O8.5, because it consistently explains the lack of an
[ArIII] line in the RCW122\_2 source, combined with the total
bolometric luminosity observed for these two sources \citep{ghosh89}.

In summary, the high spatial and low spectral resolution MIR
observations of the luminous embedded targets in the nearby high-mass
star-forming regions from our sample have led us to identify their
ionizing sources. The large uncertainties in extinction measurements
for these objects, and the lack of detection of [SIV] lines, prohibit
an accurate description of the targets. Even so, the temperature
analysis presented here, together with other diagnostics from the 
literature, clearly suggests that these are mid-late O-type objects,
three of which may evolve to become early O stars.

\begin{acknowledgements}
The authors are grateful to the referee, Dr. Lex Kaper, for constructive
and useful comments. This work was carried out under the auspices of
an European Union Marie-Curie IRSES grant (230483). This work is based
on the data obtained through the ESO VLT proposal id 075.C-0833A.1.
\end{acknowledgements}

\bibliographystyle{aa}

\end{document}